\documentclass{appolb}
\usepackage{graphicx}
\begin{document}
\newcount\eLiNe\eLiNe=\inputlineno\advance\eLiNe by -1

\title{\vspace{-20mm}
\hspace{80mm} LU-TP 12-45\\\hspace*{78mm}
December 2012\vspace*{2cm}\\
Small $x$, Saturation, and Diffraction in Collisions \\
with electrons, protons, and nuclei
\footnote{Talk presented at International Symposium on Multiparticle Dynamics,
Kielce, Poland, 16-21 September 2012.}}%
\author{G\"{o}sta Gustafson
\address{Dept. of Astronomy and Theoretical Physics,\\ 
Lund University, Lund, Sweden\\
email: gosta.gustafson@thep.lu.se}
}
\maketitle

\begin{abstract}
The Lund dipole model DIPSY is based on BFKL evolution and saturation. It can
be applied to collisions between electrons, protons, and nuclei. In this
talk I present some recent results for exclusive final states in
inelastic collisions,  a method to generate final states in diffractive
excitation, and some results for collisions with nuclei.
\end{abstract}
                          
\section*{Introduction}

The \textsc{Pythia} MC-model is the most successful 
description of inelastic reactions in DIS and $pp$ collisions. It does,
however, need input structure functions determined by data, and it also 
uses simplified assumptions about correlations and diffraction 
(with parameters retuned at different energies). Our aim is not to obtain the
most accurate description, but instead to understand underlying dynamics in 
more detail, also at the cost of lower precision. The results are obtained in
collaboration with Christoffer Flensburg and Leif L\"{o}nnblad. The outline is
the following:

1. Evolution of parton densities to small $x$

2. Correlations and fluctuations

3. Diffraction

4. Nucleus collisions

\section{Small $x$ evolution}

\subsection{Dipole cascade models}

\textbf{Mueller's dipole model} 

Mueller's dipole cascade model
\cite{Mueller:1993rr} is a formulation
of BFKL evolution in transverse coordinate space. 
Gluon radiation from the color charge in a parent quark or gluon is screened 
by the accompanying anticharge 
in the color dipole. This suppresses emissions at large transverse separation,
which corresponds to the suppression of small $k_\perp$ in BFKL.
For a dipole with transverse coordinates $(\mathbf{x},\mathbf{y})$, 
the probability per unit rapidity ($Y$) 
for emission of a gluon at transverse position $\mathbf{z}$ is given by
\begin{eqnarray}
\frac{d\mathcal{P}}{dY}=\frac{\bar{\alpha}}{2\pi}d^2\mathbf{z}
\frac{(\mathbf{x}-\mathbf{y})^2}{(\mathbf{x}-\mathbf{z})^2 (\mathbf{z}-\mathbf{y})^2},
\,\,\,\,\,\,\, \mathrm{with}\,\,\, \bar{\alpha} = \frac{3\alpha_s}{\pi}.
\label{eq:dipkernel1}
\end{eqnarray}
The dipole is split into two dipoles, which
(in the large $N_c$ limit) emit new gluons independently. The result is a
chain of dipoles, where the number of dipoles grows exponentially with $Y$.

When two cascades collide, a pair of dipoles with coordinates 
$(\mathbf{x}_i,\mathbf{y}_i)$ and $(\mathbf{x}_j,\mathbf{y}_j)$ can interact 
via gluon exchange with the probability $2f_{ij}$, where
\begin{equation}
  f_{ij} = f(\mathbf{x}_i,\mathbf{y}_i|\mathbf{x}_j,\mathbf{y}_j) =
  \frac{\alpha_s^2}{8}\biggl[\log\biggl(\frac{(\mathbf{x}_i-\mathbf{y}_j)^2
    (\mathbf{y}_i-\mathbf{x}_j)^2}
  {(\mathbf{x}_i-\mathbf{x}_j)^2(\mathbf{y}_i-\mathbf{y}_j)^2}\biggr)\biggr]^2.
\label{eq:dipamp}
\end{equation}
Summing over all dipoles in the cascades then reproduces the LL BFKL result.
%The elastic scattering amplitude is given by $T=1-\exp(-\sum f_{ij})$, and the
%cross sections are given by Eqs.~(\ref{eq:sigmat}, \ref{eq:sigmae},
%\ref{eq:sigmad}). 

\textbf{Lund cascade model DIPSY}

The Lund dipole cascade model DIPSY \cite{Avsar:2005iz,% Avsar:2006jy,
  Flensburg:2008ag} is a generalization of Mueller's model, which includes:
\vspace{2mm}

-- Important non-leading BFKL effects. The most essential ones are related to
energy conservation and the running coupling.

-- Saturation from pomeron loops in the evolution. This is not included in
Mueller's model or in the BK equation.

-- Confinement effects. Needed to satisfy the Froissart bound.

-- MC implementation DIPSY. This gives also fluctuations and correlations.

-- Applications to collisions between electrons, protons, and nuclei.
\vspace{2mm}

At high energies several dipole pairs can interact, and in the eikonal
approximation the amplitude is given by 
\begin{equation}
T=1-e^{-F},\,\,\,\,\mathrm{with}\,\,\, F=\sum f_{ij},
\label{eq:sat}
\end{equation}
where the Born amplitude
$F$ is given by summing over pairs of a dipole $i$ in the
projectile and $j$ in the target. The total and elastic cross sections are
then given by
$d \sigma_{el}/d^2 b  = T^2,\,\,\,\,\, d \sigma_{tot}/d^2 b  = 2T$.
Multiple interactions give color loops, which are related to pomeron loops.
In the DIPSY model saturation effects from color loops within the cascade
evolution are also included, which makes the result approximately independent
of the Lorentz frame used for the analysis.

The DIPSY model and the MC can be applied to collisions with electrons, protons
and nuclei. The coupling of a virtual photon to a $q\bar{q}$
dipole is determined by QED. For a proton we assume an initial wavefunction 
represented by three dipoles forming an equilateral triangle. When this is
evolved to small $x$-values, the result is not sensitive to the details of the
initial state. For a nucleus the nucleons are located according to a
Wood--Saxon distribution. Results for inclusive and (quasi)elastic
observables were presented in
Refs.~\cite{Avsar:2005iz, Flensburg:2008ag, Avsar:2007ht}.

\begin{figure}[htb]
\centerline{
\includegraphics[width = 0.33\linewidth]{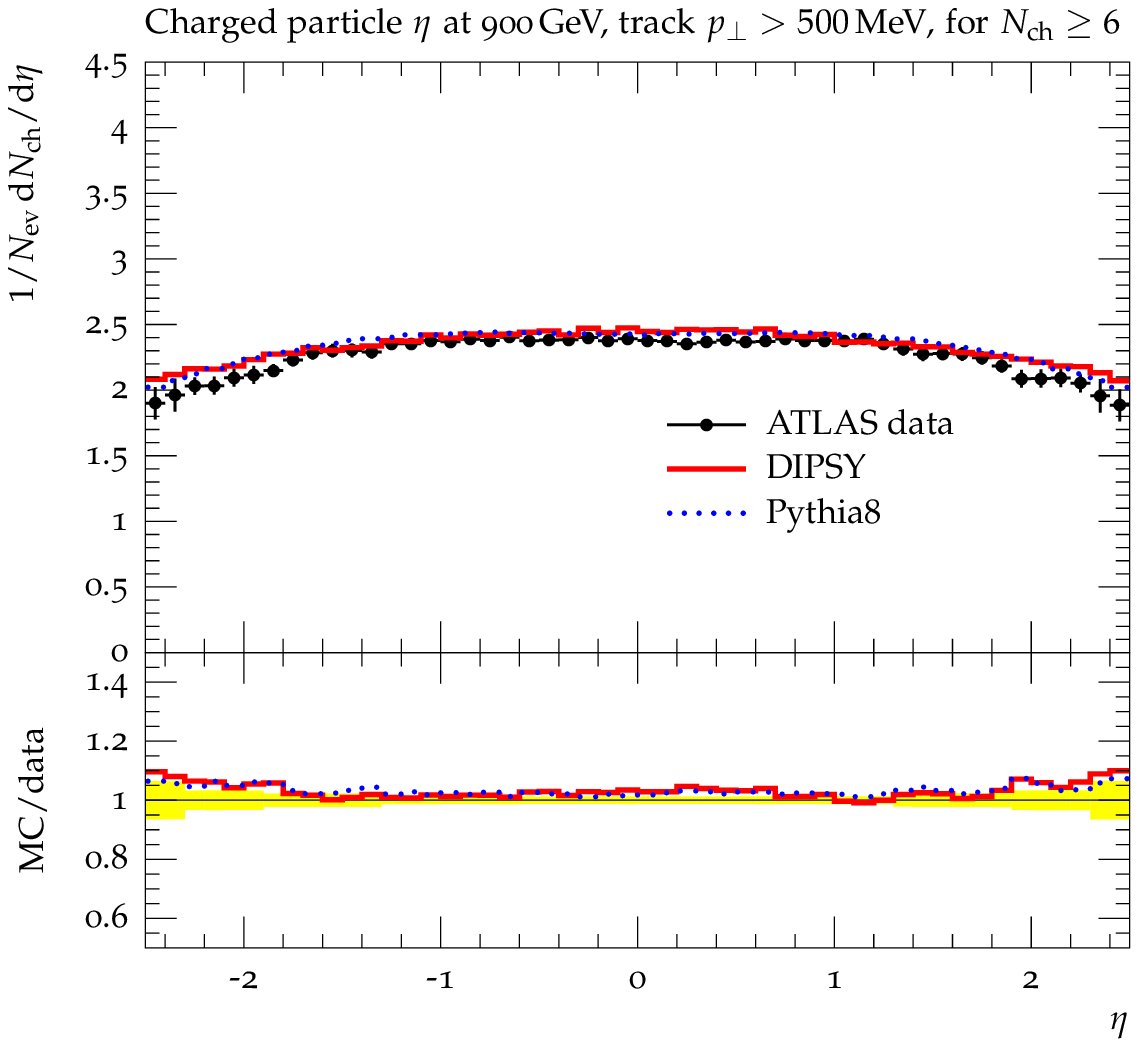}
\includegraphics[width = 0.33\linewidth]{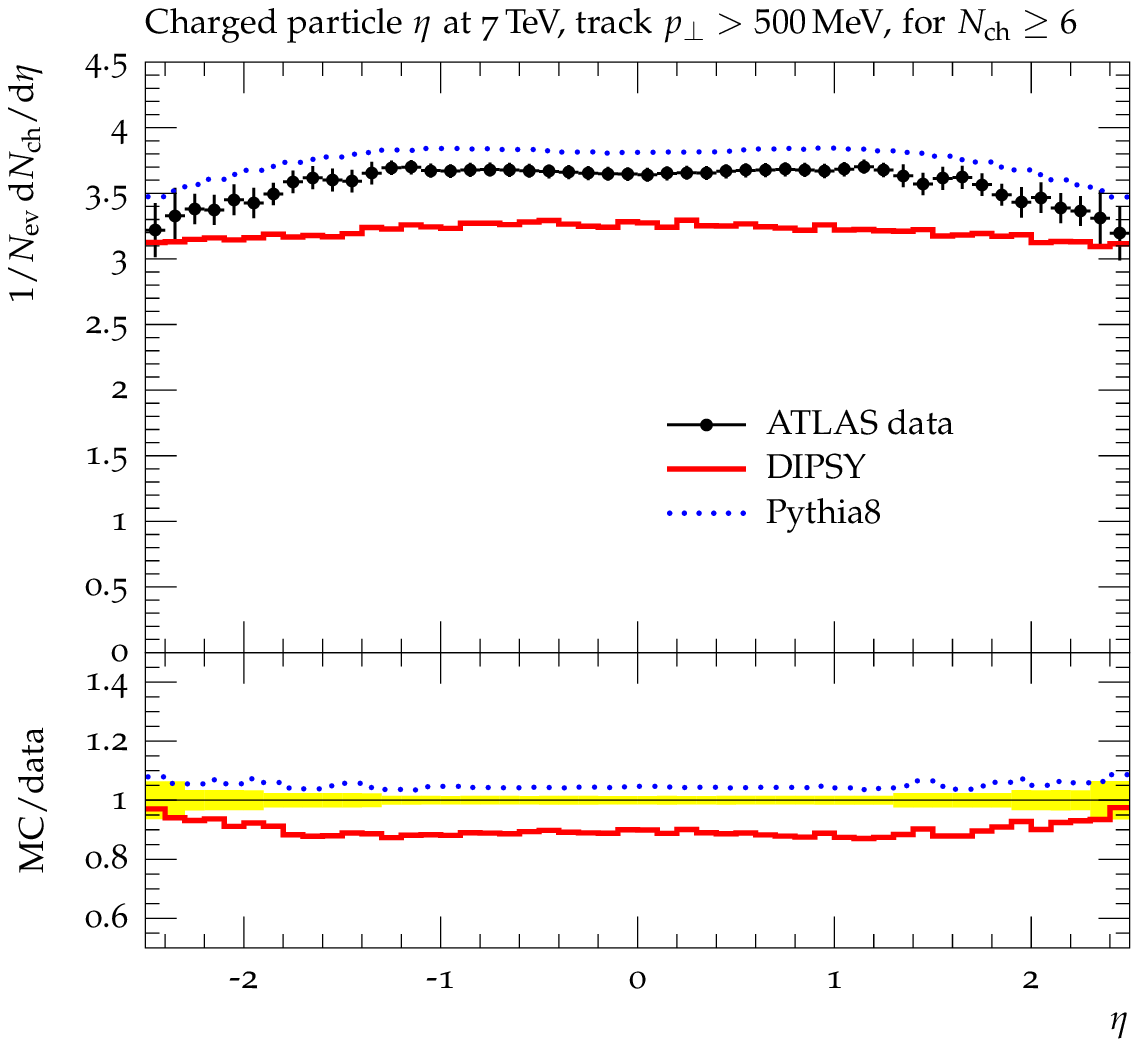}
\includegraphics[width = 0.33\linewidth]{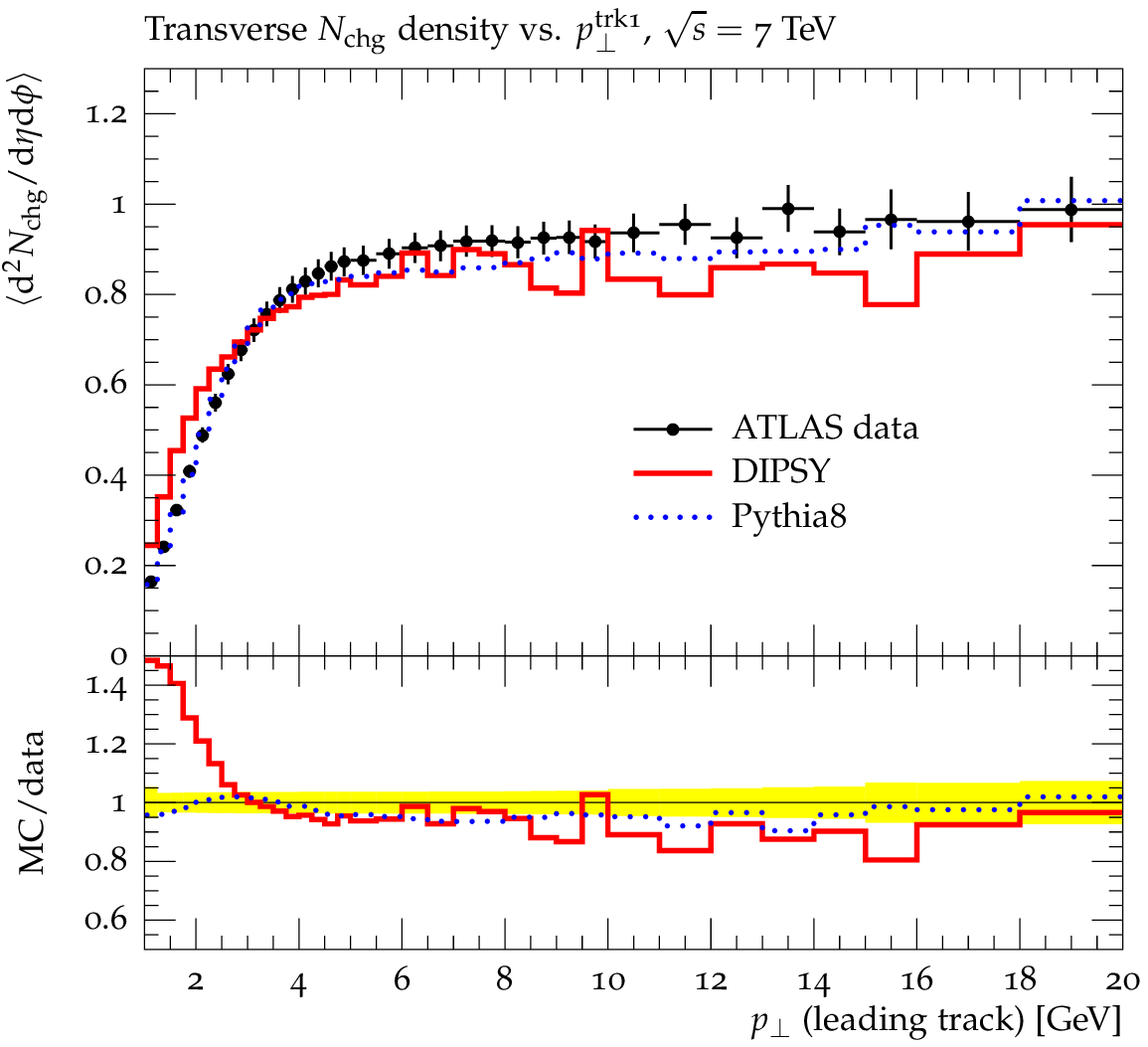}}
\caption{Min. bias $\eta$ distributions at 0.9 and 7 TeV (left and center). 
Underlying event at 7 TeV: $N_{ch}$ in transverse region \emph{vs} $p_\perp$ for a
leading particle (right). Data from ATLAS \cite{Aad:2010ac, Aad:2010ey}.}
\label{fig:ppexclusive}
\end{figure}

The model can also be used for \emph{exclusive} final states.
BFKL describes results for inclusive observables, while the CCFM
\cite{Catani:1989sg, Ciafaloni:1987ur}
model, and its generalization the LDC model \cite{Andersson:1995ju, 
Gustafson:2002kz}, can be used for a
description of exclusive states. Some results from Ref.~\cite{Flensburg:2011kk} 
are shown in Fig.~\ref{fig:ppexclusive}. 

\section{Correlations and fluctuations}

\begin{figure}[htb]
\centerline{
\includegraphics[width = 0.49\linewidth,angle=0]{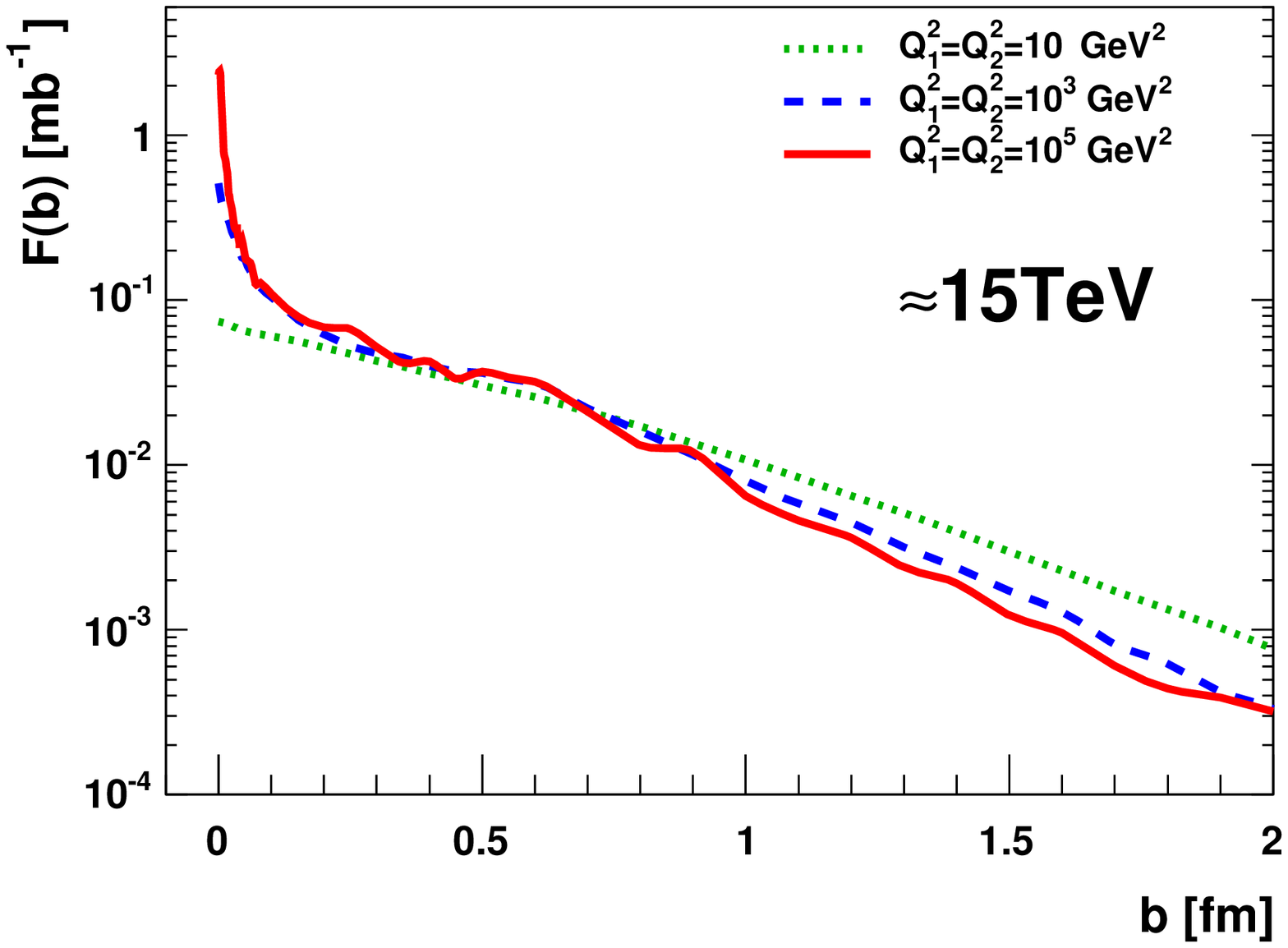}
\includegraphics[width = 0.49\linewidth]{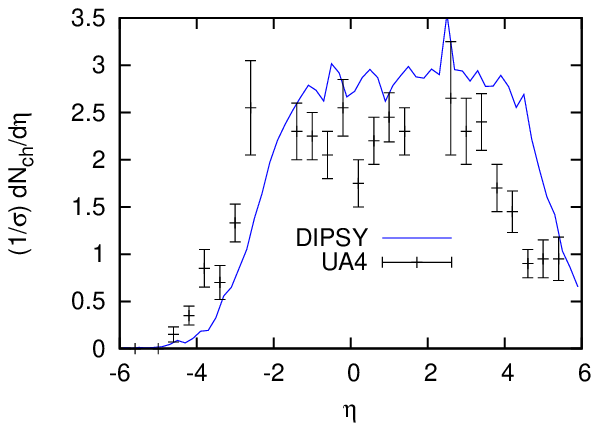}}
\caption{\emph{Left}: Two-parton impact parameter profile for proton evolution 
to $y=0$ at $\sqrt{s} =15$ TeV. \emph{Right}: $\eta$-distribution in $pp$ 
collisions at $\sqrt{s}$ = 546 GeV and 
$\langle M_X\rangle$= 140 GeV, compared to data from UA4~\cite{Bernard:1985kh}.}
\label{fig:corr}
\end{figure}

In the MC it is also possible to calculate \emph{correlations}, \emph{e.g.} 
double parton distributions relevant for multiple interactions. We define the 
double parton distribution $\Gamma$ and impact parameter profile $F$ by the 
relation
\begin{equation}
\Gamma(x_1,x_2,b;Q_1^2,Q_2^2) \equiv D(x_1, Q_1^2)\,D(x_2, Q_2^2)
\,F(b;x_1,x_2,Q_1^2,Q_2^2).
\end{equation}
Some results from Ref.~\cite{Flensburg:2011kj}
are shown in Fig.~\ref{fig:corr} \emph{left}. We note that for larger $Q^2$
hotspots develop with increased correlation at small separations $b$.

\emph{Fluctuations} are very large in QCD evolution. Taking them into account
modifies unitarization effects. When the interaction probability fluctuates
from event to event, the elastic amplitude is given by 
$\langle T\rangle =\langle 1-e^{-F}\rangle$, which 
is not equal to $1-e^{-\langle F\rangle}$. This effect
suppresses interaction for small $b$, and enhances it for larger
$b$-values \cite{Flensburg:2008ag}. Fluctuations also give odd eccentricity 
moments, such as triangular 
flow in $pp$ and $AA$ collisions (see Refs.~\cite{Avsar:2011fz,
Flensburg:2011wx}).

\section{Diffraction}

Fluctuations also cause diffractive excitations, as described in the
Good--Walker formalism. 
If the projectile has an internal structure, the mass eigenstates $\Psi_{k}$
can differ from the eigenstates of diffraction $\Phi_n$, which have different 
eigenvalues $T_n$. For an incoming proton the diffractive state is then a
different mixture of the mass eigenstates, and the inclusive cross section for
diffractive excitation is determined by the variance, 
$\langle T^2 \rangle - \langle T \rangle ^2$, of the amplitude. 
Assuming that the diffractive eigenstates are represented by parton cascades,
some results for cross sections $d\sigma/d \ln(M_X^2)$ in DIS and $pp$
scattering are presented in Ref.~\cite{Flensburg:2010kq}. The results do
reproduce experimental data from HERA and the Tevatron, and we want to
emphasize that the model is tuned only to
$\sigma_{\mathrm{tot}}$ and $\sigma_{\mathrm{el}}$, with no new free
parameter.

We also note that for $pp$ collisions the fluctuations are very much
suppressed for central collisions, when the black limit is
approached. Therefore diffractive excitation is largest for peripheral
collisions, in a circular ring expanding slowly to larger radius at higher
energy. 

\textbf{Relation Good--Walker \emph{vs} triple-pomeron}

Diffractive excitation to high masses is more traditionally described in the 
triple-regge
formalism. In Ref.~\cite{Gustafson:2012hg} it is, however, shown that the 
Good--Walker and triple-pomeron formalisms are just different ways to 
describe the same phenomenon. Such a relation was also indicated 
in Ref.~\cite{Flensburg:2010kq}, where it was demonstrated that the DIPSY 
results have the expected triple-pomeron form. Switching off the saturation
effects in the simulation, the resulting \emph{bare} pomeron
corresponds to a single pomeron pole, with an almost constant triple-pomeron
coupling. The Good--Walker formalism has then the advantage that the result
can be calculated without extra free parameters, also including saturation 
effects, which in the triple-regge formalism are described by 
``enhanced diagrams''.

\textbf{Exclusive final states in diffractive excitation}

Diffractive excitation is fundamentally a quantum effect. For exclusive final 
states different contributions interfere destructively, and therefore there is 
no probabilistic description.
Still, it is demonstrated in Ref.~\cite{Flensburg:2012zy} how the different 
contributions can be calculated within the DIPSY MC, added and squared, to
give the desired cross section. Some early applications to DIS and $pp$
collisions are presented in Ref.~\cite{Flensburg:2012zy}.
As an example Fig.~\ref{fig:corr} \emph{right} shows 
the pseudorapidity distribution in $pp$ collisions at 546 GeV.
 We note here that the $\eta$-distribution is somewhat
too hard for large $\eta$. This is due to the lack of quarks in the proton
wavefunction, and the resulting absence of a forward baryon. This has to be 
changed in
future improvements. I want to emphasize again that these results are based 
purely on fundamental QCD dynamics, with no free parameters beyond those 
tuned to the total and elastic cross sections.

\section{Nucleus collisions}

\begin{figure}[htb]
\centerline{
\includegraphics[angle=0, width = 0.48\linewidth]{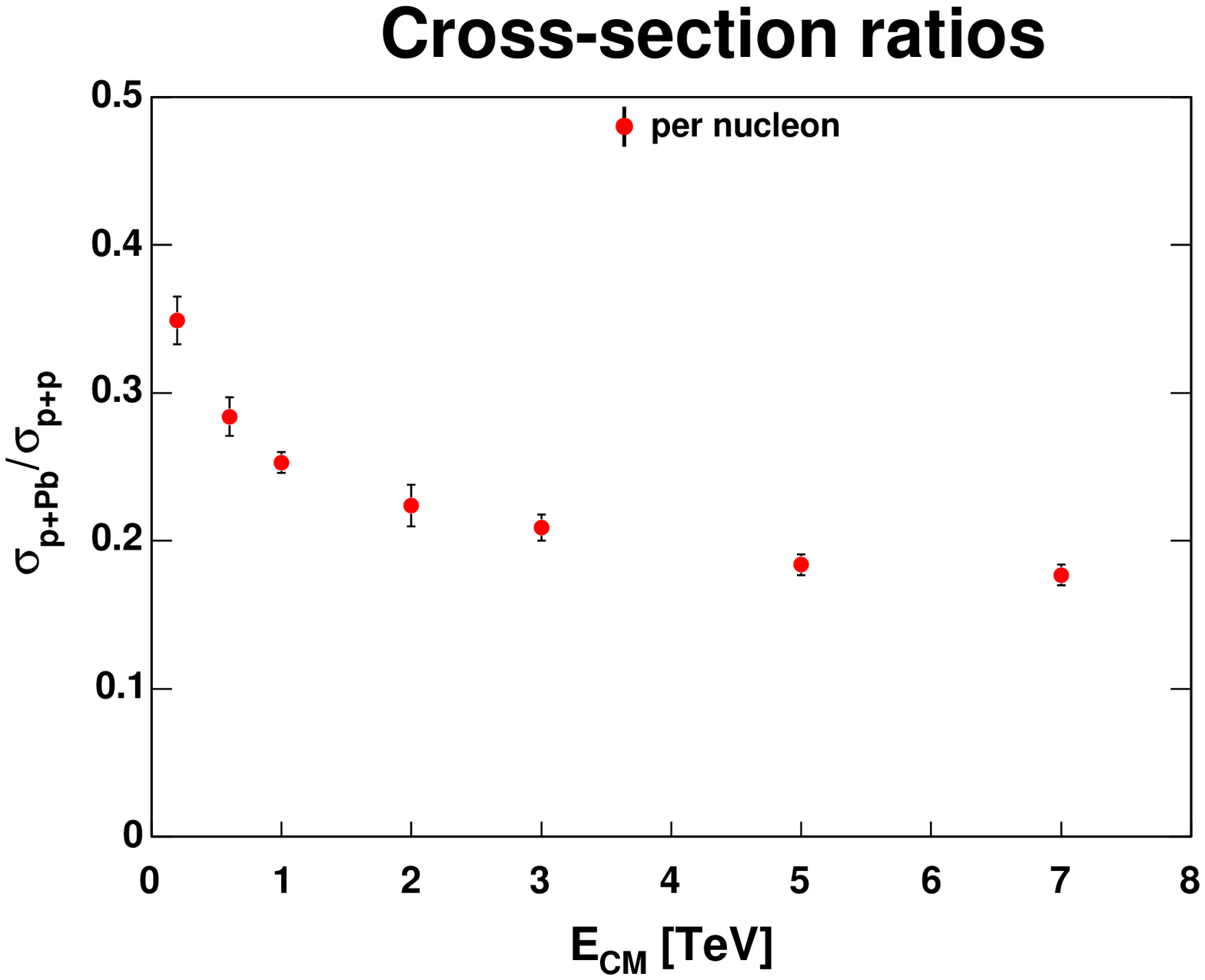}
\includegraphics[width = 0.50\linewidth,angle=0]{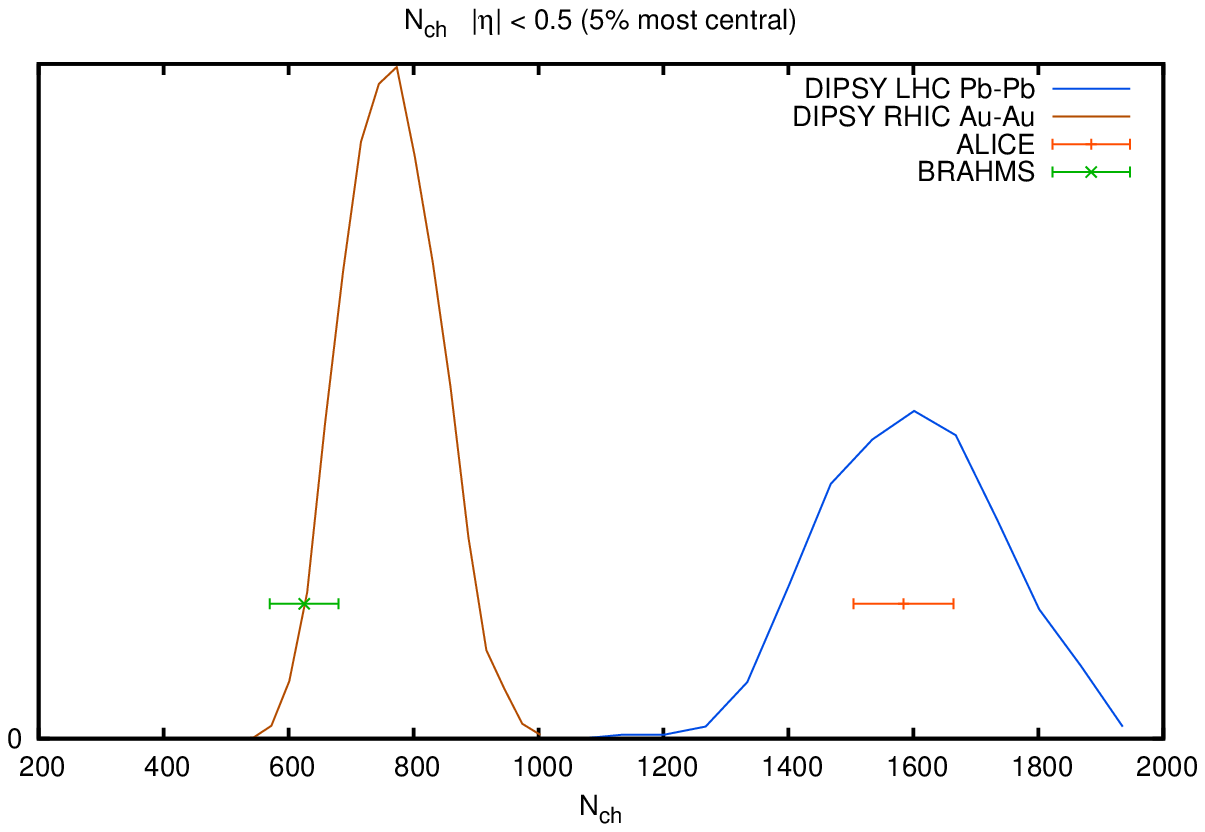}}
\caption{\emph{Left}: Shadowing effect $R=\sigma_{pA}/A \sigma_{pp}$ for $pPb$
  collisions. \emph{Right}: Charged multiplicity from toy model 
  ``thermalization'', in $|\eta|<0.5$ for central collisions at RHIC ($AuAu$) 
and LHC ($PbPb$).}
\label{fig:nucleus}
\end{figure}

The DIPSY model can also be applied to nucleus collisions. Here it gives the
full partonic picture, accounting for saturation within the cascades,
correlations and fluctuations (it gives \emph{e.g.} triangular 
flow \cite{Flensburg:2011wx}), 
and finite size effects.  As example Fig.~\ref{fig:nucleus} \emph{left} shows 
the shadowing effect in $pPb$ collisions. 

For the final state in nucleus
collisions the model gives a dense gluon soup. This state  
can be used as initial conditions in a
subsequent hydrodynamical expansion. Only adding FSR and hadronization as for
hadronic collisions, would give too many particles, but a toy model, in which
the gluons within 1 fm are allowed to interact and reconnect, might simulate the
``thermalization'' in the final state. As example Fig. \ref{fig:nucleus}
\emph{right} shows the resulting multiplicity distribution for
$|\eta|<0.5$ in central collisions at RHIC and LHC.

\section{Summary}

The DIPSY model is based on QCD dynamics for small $x$ evolution
  plus saturation. It is an attempt to understand basic dynamics, not to
  obtain optimal precision.

-- It works well for inclusive observables,

-- gives a fair description of exclusive final states,

-- describes diffraction with no extra free parameter 
    (including exclusive diffractive states),

-- includes correlations and fluctuations 
    (goes beyond the meanfield approximation in the BK eq.),

-- is also applicable to nucleus collisions, where it
accounts for saturation within
the cascades, correlations, fluctuations, 
 and finite size effects, and can give initial conditions for hydro expansion.


\begin{thebibliography}{9}

\bibitem{Mueller:1993rr}
     A.H. Mueller, 
 %    \emph{Soft gluons in the infinite momentum wave function and the
  %                BFKL pomeron},
     \emph{Nucl. Phys.} {\bf B415} (1994) 373, {\bf B437} (1995) 107.

%\cite{Avsar:2005iz}
\bibitem{Avsar:2005iz}
  E.~Avsar, G.~Gustafson, L.~L\"{o}nnblad,
  %``Energy conservation and saturation in small-x evolution,''
  JHEP {\bf 0507 } (2005)  062, {\bf 01} (2007) 012.
%  [hep-ph/0503181].

%\bibitem{Avsar:2006jy}
%     E. Avsar, G. Gustafson, and L. L\"{o}nnblad,
%     \emph{Small-x dipole evolution beyond the large-N(c) limit},
%     JHEP {\bf 01} (2007) 012 %[hep-ph/0610157].

%\cite{Flensburg:2008ag}
\bibitem{Flensburg:2008ag}
  C.~Flensburg, G.~Gustafson, L.~L\"{o}nnblad,
  %``Elastic and quasi-elastic $p p$ and $\gamma^{*} p$ scattering in the Dipole Model,''
  Eur.\ Phys.\ J.\  {\bf C60 } (2009)  233.
%  [arXiv:0807.0325 [hep-ph]].

%\cite{Avsar:2007ht}
\bibitem{Avsar:2007ht}
  E.~Avsar and G.~Gustafson,
  %``Geometric scaling and QCD dynamics in DIS,''
  JHEP {\bf 0704} (2007) 067.
%  [hep-ph/0702087 [HEP-PH]].
  %%CITATION = HEP-PH/0702087;%%

%\cite{Catani:1989sg}
\bibitem{Catani:1989sg}
  S.~Catani, F.~Fiorani, G.~Marchesini,
  %``Small x Behavior of Initial State Radiation in Perturbative QCD,''
  Nucl.\ Phys.\  {\bf B336 } (1990)  18.

%\cite{Ciafaloni:1987ur}
\bibitem{Ciafaloni:1987ur}
  M.~Ciafaloni,
  %``Coherence Effects in Initial Jets at Small q**2 / s,''
  Nucl.\ Phys.\  {\bf B296 } (1988)  49.

%\cite{Andersson:1995ju}
\bibitem{Andersson:1995ju} 
  B.~Andersson, G.~Gustafson and J.~Samuelsson,
  %``The Linked dipole chain model for DIS,''
  Nucl.\ Phys.\ B {\bf 467}, 443 (1996).
  %%CITATION = NUPHA,B467,443;%%

%\cite{Gustafson:2002kz}
\bibitem{Gustafson:2002kz} 
  G.~Gustafson, L.~Lonnblad and G.~Miu,
  %``Hadronic collisions in the linked dipole chain model,''
  Phys.\ Rev.\ D {\bf 67}, 034020 (2003).
%  [hep-ph/0209186].
  %%CITATION = HEP-PH/0209186;%%

%\cite{Flensburg:2011kk}
\bibitem{Flensburg:2011kk} 
  C.~Flensburg, G.~Gustafson and L.~L\"{o}nnblad,
  %``Inclusive and Exclusive Observables from Dipoles in High Energy Collisions,''
  JHEP {\bf 1108}, 103 (2011).
%  [arXiv:1103.4321 [hep-ph]].
  %%CITATION = ARXIV:1103.4321;%%

%\cite{Aad:2010ac}
\bibitem{Aad:2010ac} 
  G.~Aad {\it et al.}  [\textsc{Atlas} Collaboration],
  %``Charged-particle multiplicities in pp interactions measured with the ATLAS detector at the LHC,''
  New J.\ Phys.\  {\bf 13}, 053033 (2011).
 % [arXiv:1012.5104 [hep-ex]].
  %%CITATION = ARXIV:1012.5104;%%

%\cite{Aad:2010ey}
\bibitem{Aad:2010ey} 
  G.~Aad {\it et al.}  [\textsc{Atlas} Collaboration],
  %``Measurement of the top quark-pair production cross section with ATLAS in pp collisions at $\sqrt{s}=7$ TeV,''
  Eur.\ Phys.\ J.\ C {\bf 71}, 1577 (2011).
%  [arXiv:1012.1792 [hep-ex]].
  %%CITATION = ARXIV:1012.1792;%%

%\cite{Flensburg:2011kj}
\bibitem{Flensburg:2011kj}
  C.~Flensburg, G.~Gustafson, L.~L\"{o}nnblad and A.~Ster,
  %``Correlations in double parton distributions at small x,''
  JHEP {\bf 1106} (2011) 066
%  [arXiv:1103.4320 [hep-ph]].
  %%CITATION = ARXIV:1103.4320;%%

%\cite{Avsar:2011fz}
\bibitem{Avsar:2011fz}
  E.~Avsar, Y.~Hatta, C.~Flensburg, J.~Y.~Ollitrault and T.~Ueda,
  %``Eccentricity and elliptic flow in pp collisions at the LHC,''
  J.\ Phys.\ G {\bf 38} (2011) 124053.
%  [arXiv:1106.4356 [hep-ph]].
  %%CITATION = ARXIV:1106.4356;%%

%\cite{Flensburg:2011wx}
\bibitem{Flensburg:2011wx}
  C.~Flensburg,
  %``Correlations and Fluctuations in the Initial State of high energy Heavy
  %Ion Collisions,''
    arXiv:1108.4862 [nucl-th].

%\cite{Flensburg:2010kq}
\bibitem{Flensburg:2010kq}
  C.~Flensburg and G.~Gustafson,
  %``Fluctuations, Saturation, and Diffractive Excitation in High Energy Collisions,''
  JHEP {\bf 1010} (2010) 014.
%  [arXiv:1004.5502 [hep-ph]].
  %%CITATION = ARXIV:1004.5502;%%

%\cite{Gustafson:2012hg}
\bibitem{Gustafson:2012hg}
  G.~Gustafson,
  %``The Relation between the Good-Walker and Triple-Regge Formalisms for Diffractive Excitation,''
  arXiv:1206.1733 [hep-ph]. To be published in Phys. Lett. B.
  %%CITATION = ARXIV:1206.1733;%%

%\cite{Flensburg:2012zy}
\bibitem{Flensburg:2012zy} 
  C.~Flensburg, G.~Gustafson and L.~L\"{o}nnblad,
  %``Exclusive final states in diffractive excitation,''
  arXiv:1210.2407 [hep-ph].
  %%CITATION = ARXIV:1210.2407;%%

%\cite{Bernard:1985kh}
\bibitem{Bernard:1985kh}
  D.~Bernard, 1 {\it et al.}  [UA4 Collaboration],
  %``Pseudorapidity Distribution Of Charged Particles In Diffraction Dissociation Events At The Cern Sps Collider,''
  Phys.\ Lett.\ B {\bf 166} (1986) 459.
  %%CITATION = PHLTA,B166,459;%%


\end{thebibliography}
\end{document}